\documentstyle[12pt]{article}
\newcommand{\ti}{\tilde}

\def\t{\tau}

\def\bba{\begin{array}}
\def\eea{\end{array}}

\def\ti{\tilde}

\def\bb{\begin{equation}}
\def\ee{\end{equation}}

\begin{document}
\large

%---------------------------------------------------------- TITLE --------

\bigskip
\centerline{\LARGE Integrable boundary conditions}
\centerline{\LARGE  for nonlinear
lattices\footnote{Supported by RFBR, Grant \# 98-01-00576}}

\bigskip
\bigskip
\centerline{I.T.Habibullin and A.N.Vil'danov}
\bigskip

\abstract{Integrable boundary conditions in 1+1 and 2+1
dimensions are discussed from the higher symmetries point of view. Boundary
conditions consistent with the discrete Landau-Lifshitz model and infinite
2D Toda lattice are represented.}

%============== SECTION 1 ===============================================
\section{Introduction}

The inverse scattering transform method is a powerful tool for solving
the Cauchy problem for nonlinear integrable equations. However the
method is not sufficiently effective in application to the initial
boundary value problems on a half line or a finite interval when both
the boundary condition and the initial data are chosen arbitrary. At
the same time there is a special kind of boundary conditions, called
integrable, which are completely consistent with the integrability
property of the equation. Such that the inverse scattering method is
effectively applied to the initial boundary value problem when these
boundary conditions are imposed. Several definitions of integrable
boundary conditions are known in the literature (see, for instances,
\cite{skl,ggh}). In essence they are more or less equivalent. We will
concentrate ourselves on one of them based on the symmetry approach.

During the last decade the classes of integrable boundary
conditions have been studied for a large number of
physically interesting equations in dimension 1+1 like the
sine-Gordon equation \cite{skl,bt,gz}, affine Toda
lattices \cite{cor}, the KdV equation \cite{aggh} etc.

In the dimensions higher than 1+1 the problem is still less studied. The
main difficulty appearing in multi-dimensionality is connected with the
so-called non-local dynamical variables  higher symmetries and conserving
quantities depend on. Recently \cite{gh} it has been shown that the
symmetry approach can effectively be used to find integrable boundary
conditions in 1+2 dimensions.

In the present paper a boundary condition is found consistent with the Toda
lattice by making use of the symmetry method. The boundary 2D lattice found
reduced into the well-known boundary affine Toda field equations \cite{cor}
by imposing the periodicity and similar other closure constraints.

Another model dealt with is the discrete version of the famous
Landau-Lifshitz equation. Some integrable boundary conditions are
represented for this chain. The boundary conditions found are
differential constraints consistent with the nearest symmetry of the
continuous L-L equation. An interesting fact is that under this
constraint the symmetry mentioned turns into the Krichever-Novikov
equation.

%============== SECTION 1 ==============================================

\section{Boundary condition for the discrete L-L model}

Let us consider an integrable lattice of the form
\bb
u_x=f(u(n-1),u(n),u(n+1)),\quad -\infty<n<+\infty,
                                                               \label{ux}
\ee
where $u=u(n,x)$ is the unknown function and  lower index
denotes the derivative. Impose the following boundary
condition at the point $n=k$
\bb
u(k)=F(u(k+1),...,u(k+m))\,.
                                                               \label{uk}
\ee
Under this constraint the lattice (\ref{ux}) turns into a
semi-infinite one, defined for all $n\geq k+1$. Let us given one more
lattice of similar form
\bb
u_t(n)=g(u(n-s),u(n-s+1),...,u(n+s))\,,
                                                               \label{ut}
\ee
which is a symmetry of the lattice (\ref{ux}), i.e. two flows defined
by chains (\ref{ux}) and (\ref{ut}) commute: $(u_x(n))_t=(u_t(n))_x$.
It is clear that the constraint (\ref{ux}) and its differential
consequences reduce the infinite chain (\ref{ut}) to a semi-infinite
chain defined for $n\geq k+1$. The boundary condition (\ref{uk}) is
called consistent with the symmetry (\ref{ut}), if these two
semi-infinite lattices obtained are commuting.

If the boundary condition is consistent with an infinite set of symmetries
of the lattice then it is called consistent with the integrability.

The commutativity of two semi-infinite lattices mentioned above might be
checked directly but in some cases it is more convenient to utilize to this
purpose the so called associated system. Let us rewrite the pair of
equations (\ref{ux}), (\ref{ut}) as a system of partial differential
equations. To this end introduce another set of dynamical variables,
serving these chains and consisting of $u=u(k), v=u(k+1)$ and their
$x-$derivatives $u_x,v_x,u_{xx},v_{xx},\dots$. One expresses $u(n)$ for
$n<k$ and $n>k+1$ through the new variables by using the equation
(\ref{ux}) and its differential consequences. Substitute now the
expressions obtained instead of the variables $u(n)$ into the chain
(\ref{ut}). As a result one gets a system of the form
\begin{eqnarray}
   u_t&=&g_1(u,v,u_x,v_x,...,u_s), \nonumber \\
   v_t&=&g_2(u,v,u_x,v_x,...,v_s),    \label{asy}
\end{eqnarray}
where $\displaystyle{u_s={\partial^s u\over\partial x^s}}$ ,
$\displaystyle{v_s={\partial^s v\over\partial x^s}}$. The boundary
condition (\ref{uk}) reads as
\bb
 u=G(v,u_x,v_x,...,u_m).    \label{cons}
\ee

The following important statement is a consequence of
these transforms \cite{ggh}.

{\bf Proposition.} The boundary condition (\ref{uk}) is
consistent with the symmetry (\ref{ut}) if and only if the
constraint (\ref{cons}) is consistent with the associated
system (\ref{asy}).

It is clear that among the symmetries consistent with the integrable
boundary condition one can find that of the smallest order. We refer to
such a symmetry as a trial one. Our hypothesis, approved by numerous
examples, claims that for the integrable equation given all trial
symmetries connected with the certain type boundary conditions have one and
the same order and this order can be pointed out a priori by rather simple
preliminary analysis.

Let us consider an illustrative example. The following
integrable discrete analogue of the well known
Landau-Lifshits equation was found in \cite{shayam} years
ago:
\bb u_{nx}=v_n,\quad v_{nx}=-f_n,        \label{ll}
\ee
where the function $f_n$ is defined as
$$ f_n=(v_n^2+P(u_n))\left(\frac{1}{u_{n+1}-u_{n}}-
\frac{1}{u_{n}-u_{n-1}}\right)+\frac{P'(u_n)}{2}\,,$$
and $P(u)=au^4+bu^3+cu^2+du+e$ is an arbitrary polynomial
of forth order with constant coefficients. The lattice
(\ref{ll}) admits higher symmetries, the simplest one is
of the form
\bb u_{nt}=h_n,\quad v_{nt}=D_x(h_n),    \label{sl1}
\ee
where
$$ h_n=(v_n^2+P(u_n))\left(\frac{1}{u_{n+1}-u_{n}}+
\frac{1}{u_{n}-u_{n-1}}\right)\,.$$
Let us introduce new variables $q=u_{-1}$ and $u=u_0$ and then by
excluding the explicit $n$-dependence, as it was discussed above,
bring the symmetry (\ref{sl1}) to the form of associated system
\begin{eqnarray}
u_t&=&u_{xx}-2\frac{u_x^2+P(u)}{u-q}+\frac{P'(u)}{2},\nonumber\\
-q_t&=&q_{xx}+2\frac{q_x^2+P(q)}{u-q}+\frac{P'(q)}{2},  \label{cll}
\end{eqnarray}
which is an integrable generalization of the Landau-Lifshits model. In the
particular case, if the polynomial $P(u)$ subject to the additional
constraint $P(u)=au^4+cu^2+a$ then the system (\ref{cll}) coincides with
the L-L model taken under the stereographic projection. The next symmetry
of the chain (\ref{ll}) in terms of the variables $q,u$ is at the same time
a symmetry of the system (\ref{cll}). We take it in the form of the
associated system
\begin{eqnarray}
&u_{\tau}=\displaystyle{ 2u_{xxx}+u_xP''(u)-\frac{12u_x}{u-q}(u_{xx}+
\frac{P'(u)}{2})+\frac{12u_x}{(u-q)^2}(u_{x}^2+P(u)),}\nonumber\\
&q_{\tau}=\displaystyle{ 2q_{xxx}+q_xP''(q)+\frac{12q_x}{u-q}(q_{xx}+
\frac{P'(q)}{2})+\frac{12q_x}{(u-q)^2}(q_{x}^2+P(q)).}\label{sl2}
\end{eqnarray}
To find the boundary condition of the form
\bb
u_{-1}=F(u_0)
\label{bcll}
\ee
consistent with the integrability property of the
Landau-Lifshits chain one has to answer the question when
the constraint of the form $q=H(u)$ is compatible with the
trial symmetry (\ref{sl2}). Only three choices are
possible: $$1)\,q=c, \quad 2)\,q=-u+c,\quad3)\,
(c_1q+c_2)(c_1u+c_2)=-1.
$$
Here $c,$ $c_1,$ and $c_2$ are arbitrary constants.

%============== SECTION 2 ==============================================

\section{Higher dimensions}

In dimensions higher than 1+1 the phenomenon of integrable boundary
conditions is less studied. The classical generalized Toda lattices
corresponding to infinite series of Lie algebras of finite growth can be
interpreted as finite reductions of infinite 2D Toda lattice
\begin{equation}
q_{xy}(n)=e^{q(n+1)-q(n)}-e^{q(n)-q(n-1)},
\label{tl}
\end{equation}
with integrable cutting off conditions  at two fixed
points $n=0,$ $n=N.$ In \cite{gh} the question was
examined when the boundary condition of the form
\begin{equation}
q(1)=F(q(0),q_x(0),q_y(0),q(-1))
                                                                \label{bc}
\end{equation}
is consistent with the integrability property of the Toda lattice
(\ref{tl}). In higher dimensions the main obstacle is connected with the
non-local variables which higher symmetries depend on. For instance, the
following two equations
\begin{eqnarray}
\displaystyle{q_{t_1}}(n)=b_1(n)+b_1(n-1)+q_x(n)^2                             \label{tos}
\end{eqnarray}
and
\begin{eqnarray}
q_{t}(n)&=&b_2(n-2)+b_2(n-1)+b_2(n)+ b_1(n)[2q_x(n)+q_x(n+1)]\nonumber\\
 & &+b_1(n-1)[2q_x(n)+q_x(n-1)]+q_x(n)^3                        \label{ts}
\end{eqnarray}
are two symmetries of the Toda lattice (\ref{tl}).  They
depend on two non-localities $b_1(n)$ and $b_2(n)$ which
are introduced as solutions to the equations \cite{shab}:
\begin{eqnarray}
q_{xx}(n)&=&b_1(n)-b_1(n-1)\,,\label{none1}\\
b_{1,y}(n)&=&c(n)[q_x(n+1)-q_x(n)]\,,\label{none2}\\
b_{1,x}(n)&=&b_1(n)[q_x(n+1)-q_x(n)]+b_2(n)-b_2(n-1)\,,\label{none3}\\
b_{2,y}(n)&=&c(n)b_1(n+1)-c(n+1)b_1(n)\,,\label{none4}
\end{eqnarray}
where $c(n)$ is described by the equation $c(n)=e^{q(n+1)-q(n)}$.
Other non-local variables $b_j,$ $j>1$ satisfy the similar equations:
\begin{eqnarray}
b_{j,x}(n)&=&b_j(n)[q_x(n+j)-q_x(n)]+b_{j+1}(n)-b_{j+1}(n-1)\,,\nonumber\\
b_{j+1,y}(n)&=&c(n)b_{j}(n+1)-c(n+j)b_j(n)\,.\nonumber
\end{eqnarray}
Let us pass from the standard set of local dynamical variables

\noindent
$q(n),q_x(n),
q_{xx}(n),...,q_y(n),q_{yy}(n),...$ for all $n=0,\pm 1,\pm 2,...$

\noindent
to the set consisting of variables $u,v$ and their all
$x$- and $y$-derivatives by setting $u=e^{q(1)},\,$
$v=e^{-q(0)}\,.$ For example, $q(-1)$ may be expressed
from the Toda equation written in the form $
e^{-q(-1)}=e^{q(1)-2q(0)}-q_{xy}(0)e^{-q(0)}\,,$ and so
on. In terms of these new variables the symmetries
(\ref{tos}) and (\ref{ts}) become (see, also \cite{shab})
\bb
\displaystyle{u_{t_1}=u_{xx}+2ru\quad, v_{t_1}=-v_{xx}-2rv}\,,
\label{symme}
\ee
\begin{eqnarray}
\begin{array}{rcl}
u_{t}&=&u_{xxx}+3ru_x+3su\,,\\
v_{t}&=&v_{xxx}+3rv_x-3sv+3r_xv\,,
\end{array}
\label{symm}
\end{eqnarray}
where non-localities $r=b_1(0)$ and $s=b_2(0)+r(\log u)_x$
obviously satisfy the equations $ r_y=(uv)_x,\quad
s_y=(u_xv)_x\,.$ The boundary condition (\ref{bc}) takes
the form
\begin{eqnarray}
u=\tilde F(v,v_x,v_y,v_{xy})\,.      \label{newbc}
\end{eqnarray}

A very useful consequence of the change of variables is
the following statement:

\noindent
{\bf Proposition.} The boundary condition (\ref{bc}) is
compatible with the symmetry (\ref{ts}) (or (\ref{tos}))
if and only if the constraint (\ref{newbc}) is consistent
with the system (\ref{symm}) (or (\ref{symme})).

\noindent
{\bf Remark.} The connection between the "new" and "old"
dynamical variables has some singularities at points $u=0$
and $v=0$ because of the formulae $s=b_2(0)+r(\log u)_x$
and $s=b_2(-1)-r(\log v)_x+b_{1,x}(0)$. That is why these
cases should be checked directly without passing to the
associated system.

In \cite{gh} two constraints has been classified
\begin{equation}
u=\ti F(v,v_x,v_y,v_{xy})\,,
\label{ex1}
\end{equation}
and
\begin{equation}
u_y=\ti G(u,v,v_x)\,.
\label{ex2}
\end{equation}
In both cases the equation (\ref{symm}) was taken as the trial symmetry
because the other one (\ref{symme}) does not admit any constraint of the
above forms, except two degenerate ones $u=0$ and $v=0$ (see Remark above),
for it is skew-symmetric in the highest order derivatives.

The consistency condition with the $t$-dynamics allows one to extract the
following five choices for the function $\ti F$, here we give also the
additional constraint the non-localities have to satisfy to:

\noindent
{\bf (i)} $u=0\,,\;s_y=0\,;$

\noindent
{\bf (ii)} $ u=a\,,\; s=0\,;$

\noindent
{\bf (iii)} $ \displaystyle{u=av\,,\;s={1\over2}r_x\,;}$

\noindent
{\bf (iv)} $\displaystyle{ u={v_{xy}\over (a-v^2)}+{vv_xv_y \over
(a-v^2)^2}\,, \quad s=r_x-{v_xv_{xx} \over a-v^2}-{vv_x^3 \over
(a-v^2)^2}};$

\noindent
here $a$ is an arbitrary constant,

\noindent
{\bf (v)} $\displaystyle{ u=-{v_{xy}\over v^2}+{v_xv_y \over v^3}\,,
\quad r=-{v_{xx} \over v}+{v_x^2 \over v^2}+b\,}$

\noindent
and $b$ is an arbitrary function of $x$.

Rewritten in terms of the basic variables these choices
read:

\noindent
{\bf (1)} $\displaystyle{e^{q(1)}=0\,,\;b_{2,y}(0)=0}\,;$

\noindent
{\bf (2)} $\displaystyle{q(1)=\mbox{const.}\,,\;b_2(0)=0}\,;$

\noindent
{\bf (3)} $\displaystyle{q(1)=-q(0)+\mbox{const.}\,,\;b_2(0)=
{b_{1,x}(0) \over 2} +
b_1(0)q_x(0)}\,;$

\noindent
{\bf (4)} $\displaystyle{ae^{q(1)}=e^{-q(-1)}+{aq_x(0)q_y(0) \over ae^{q(0)}-
e^{-q(0)}}}$\,,

\noindent
\hspace{.5cm}
$\displaystyle{b_2(0)=b_{1,x}(0)-b_1(0)q_x(1)+{aq_x(0)^3 \over  (ae^{q(0)}-
e^{-q(0)})^2}-{q_x(0)q_{xx}(0)e^{-q(0)} \over  ae^{q(0)}-e^{-q(0)}}}\,;$

\noindent
{\bf (5)} $\displaystyle{e^{-q(-1)}=0\,,\;b_{1,y}(-1)=0}.$

All these boundary conditions are known as integrable \cite{mop}.
Among constraints of the form (\ref{ex2}) the only is consistent with
the trial symmetry (\ref{symm}): $\ti G
= av_x,$ where $a\neq0$ is an arbitrary constant (if $a=0$ then
the constraint $u_y=0$ is reduced by integration to the form $u=const$).
Here non-localities should satisfy the constraint
$\displaystyle{s=u_x-{uu_x\over a}}$.

Turning back to the original variables yields the following boundary
condition
\begin{eqnarray}
q_y(1)&=&-ae^{-q(1)-q(0)}q_x(0)\,,              \label{bc41}
\end{eqnarray}
which is not reduced to any of standard ones. So the following system
of hyperbolic equations
\begin{eqnarray}
q_x(0)&=&-a_1e^{q(1)+q(0)}q_y(1)\,,              \nonumber\\
q_{xy}(j)&=&e^{q(j+1)-q(j)}-e^{q(j)-q(j-1)}\,, 1\geq j\geq N, \label{tl1}\\
q_y(N+1)&=&-a_Ne^{-q(N+1)-q(N)}q_x(N)\,,              \nonumber
\end{eqnarray}
is an integrable reduction of the 2D Toda lattice.
Similarly one can reduce the Toda lattice (\ref{tl}) by
imposing different kind closure constraints at the ends:
\begin{eqnarray}
q_x(0)&=&-a_1e^{q(1)+q(0)}q_y(1)\,,              \nonumber\\
q_{xy}(j)&=&e^{q(j+1)-q(j)}-e^{q(j)-q(j-1)}\,, 1\geq j\geq N,\label{tl2}\\
q(N+1)&=&\ti F(q(N),q_x(N),q_y(N),q(N-1))\,,              \nonumber
\end{eqnarray}
where $\ti F$ is one of the boundary conditions (1)-(5) above. Undoubtedly
the systems (\ref{tl1}), (\ref{tl2}) are integrable, but it is not clear
what is the algebraic structure they are related to.

%============== SECTION 3 ==============================================

\section{Higher dimensions -- boundary conditions of
the other kind}

Now let us represent the Toda lattice as an infinite system of hyperbolic
equations
\bb
q_{tt}(n)-q_{zz}(n)=e^{q(n+1)-q(n)}-e^{q(n)-q(n-1)},
\quad -\infty <
n<+\infty,
\label{tl3}
\ee
where the new independent variables are introduced as follows $t=x+y$
$z=y-x.$ Consider this system on the half-line $z>0$ imposing along
the border $z=0$ a boundary condition of the form
\begin{eqnarray}
q_{z}(n)&=&H_n(q(n+1),q(n),q(n-1)),\quad -\infty <n<+\infty,
\label{bc3}
\end{eqnarray}
The problem is now to extract from the class of boundary conditions
(\ref{bc3}) those consistent with the integrability. To solve the
problem we will use the symmetry approach. First choose up the trial
symmetry and rewrite it as an associated system. Notice that under the
reflection type transformation $x\rightarrow y,$ $y\rightarrow x$ the
symmetries of the Toda lattice turn info symmetries, for the lattice
itself is invariant under this transformation. So the symmetry
(\ref{tos}), (\ref{none1}) produces a new one:
$\displaystyle{q_{t_2}(n)}=p_1(n)+p_1(n-1)+q_y(n)^2\,,$
$p_1(n)=p_1(n-1)+q_{yy}(n)\,.$ The sum of this symmetry and its
counterpart (\ref{tos}), (\ref{none1}) is again a symmetry but now it
is invariant under the $x,y$`reflection $x\rightarrow y,$
$y\rightarrow x$:
\begin{eqnarray}
q_{\t}(n)&=&h(n)+h(n-1)+q_x(n)^2+q_y(n)^2\,,\nonumber\\
h(n)&=&h(n-1)+q_{xx}(n)+q_{yy}(n)\,,  \label{ds}
\end{eqnarray}
here $h(n)=p_1(n)+b_1(n).$ Notice that the symmetry (\ref{ds}) can be
reduced to the famous Davey-Stewardson equation. In terms of the new
independent variables $t,z$ it reads
\begin{eqnarray}
q_{\t}(n)&=&4q_{tt}(n)-2c(n)+2c(n-1)+2h(n-1)+2q_t(n)^2+2q_z(n)^2\,,\nonumber\\
h(n)&=&h(n-1)+4q_{tt}(n)-2c(n)+2c(n-1)\,.  \label{ds1}
\end{eqnarray}
This symmetry will be taken as a trial one. The dynamical set of
variables serving the system (\ref{ds1}) consists of the local
variables $q(n),q_z(n)$  and their $t$-derivatives for all $n$ and
non-local variables $h(n)$ and their $z$-derivatives. Let us put
$v(n)=q_z(n)$, $g(n)=h_z(n)$ and represent the symmetry (\ref{ds1}) as
an associated system
\begin{eqnarray}
q_{\t}(n)&=&4q_{tt}(n)-2c(n)+2c(n-1)+2h(n-1)+2q_t(n)^2+2v(n)^2\,,\nonumber\\
v_{\t}(n)&=&4v_{tt}-2c(n)(v(n+1)-v(n))+2c(n-1)(v(n)-v(n-1))+\nonumber\\
 & &+2g(n-1)+4q_t(n)v_t(n)+4v(n)(q_{tt}(n)-c(n)+c(n-1))\,,\nonumber\\
h(n)&=&h(n-1)+4q_{tt}(n)-2c(n)+2c(n-1)\,,\label{ads}  \\
g(n)&=&g(n-1)+4v_{tt}-2c(n)(v(n+1)-v(n))+\nonumber\\
&&+2c(n-1)(v(n)-v(n-1))\,.\nonumber
\end{eqnarray}
The boundary condition (\ref{bc3}) for the Toda lattice (\ref{tl3}) is
consistent with the trial symmetry (\ref{ds1}) if and only if the
differential constraint
\bb
v(n)=H_n(q(n-1),q(n),q(n+1))\,.
                                                               \label{abs}
\ee
is consistent with the system of equations (\ref{ads}). Direct
computations show that only choice of $H$ is (remind that $v=q_z$)
\bb
q_{z}(n)|_{z=0}=\displaystyle{c_ne^{\frac{q(n+1)-q(n)}{2}}-
c_{n-1}e^{\frac{q(n)-q(n-1)}{2}}\,}, \label{ans}
\ee
where $c_n^2=1$ for all $n$, either $c_n=0$ for all $n$. Combining the
periodicity closure constraint $q(n)=q(n+N)$ with the boundary
condition (\ref{ans}) one gets a finite system of hyperbolic equations
on a half-plane
\begin{eqnarray}
&&q_{tt}(n)-q_{zz}(n)=e^{q(n+1)-q(n)}-e^{q(n)-q(n-1)},
\quad 0 <
z<+\infty,\nonumber\\ &&q_{z}(n)|_{z=0}=\displaystyle{
c_ne^{\frac{q(n+1)-q(n)}{2}}-c_{n-1}e^{\frac{q(n)-q(n-1)}{2}}},
\label{ans2}\\
&& q(0)=q(N),\quad q(N+1)=q(1).\nonumber
\end{eqnarray}
found recently by E.Corrigan et al. (see \cite{cor}). Imposing the
closure conditions of the form (1)-(5) in addition to the boundary
condition (\ref{ans}) leads again to integrable boundary value problem
from \cite{cor}. This kind problems for finite Toda systems have
beautiful interpretation in the field theory.

%---------------------------------------------------------- REFERENCES ---

\end{document}